\documentclass[preprint]{aastex}
\shorttitle{Balmer and Paschen jump temperature determinations 
in  low-metallicity emission-line galaxies}
\shortauthors{N. G. Guseva et al.}

\begin{document}

\title{Balmer and Paschen jump temperature determinations in  
low-metallicity emission-line galaxies}

\author{Natalia G. Guseva}
\affil{Main Astronomical Observatory, National Academy of Sciences of Ukraine,
03680, Kyiv, Ukraine}
\email{guseva@mao.kiev.ua}

\author{Yuri I. Izotov}
\affil{Main Astronomical Observatory, National Academy of Sciences of Ukraine, 
03680, Kyiv, Ukraine}
\email{izotov@mao.kiev.ua}

\and 

\author{Trinh X. Thuan}
\affil{Astronomy Department, University of Virginia,
    Charlottesville, VA 22903}
\email{txt@virginia.edu}

\begin{abstract}
We have used the Balmer and Paschen jumps to determine the temperatures 
of the H$^+$ zones of a total sample of 47 H {\sc ii} regions. 
The Balmer jump was used on MMT spectrophotometric data
of 22 low-metallicity H {\sc ii} regions in 
18 blue compact dwarf (BCD) galaxies and of one H {\sc ii} region in the spiral
galaxy M101. The Paschen jump was used  
on spectra of 24 H {\sc ii} emission-line galaxies selected  
from the Data Release 3 of the Sloan 
Digital Sky Survey (SDSS). 
To derive the temperatures, we have used a 
Monte Carlo technique varying the electron temperature in the H$^+$ zone,
the extinction of the ionized gas and that of the stellar population,
the relative contribution of the ionized gas to the total emission and 
the star formation history  
to fit the spectral energy distribution of the galaxies.
For the MMT spectra, the fit was done  
in the wavelength range of 3200 -- 5200\AA\  
which includes the Balmer discontinuity, and for the SDSS spectra, 
in the wavelength range
of 3900 -- 9200\AA\ which includes the Paschen 
discontinuity. We find for our sample of H {\sc ii} regions 
that the
temperatures of the O$^{2+}$ zones determined from the nebular to auroral line 
intensity ratio of doubly ionized 
oxygen [O {\sc iii}] $\lambda$(4959+5007)/$\lambda$4363
do not differ, in a statistical sense, from the 
temperatures of the H$^+$ zones determined from fitting the Balmer 
and Paschen jumps and the SEDs. We cannot rule out small temperature 
differences of the order of 3--5\%. 

\end{abstract} 

\section {Introduction}
\label{intro}

   The knowledge of the temperature structure of an H {\sc ii} region is
essential for the determination of its element abundances.  Heavy element 
abundances are typically derived from fluxes
of collisionally excited forbidden lines in the optical range and they depend  
very sensitively (exponentially) 
on the adopted electron temperature. Since different ions
reside in different parts of the H {\sc ii} region with different electron
temperatures, an appropriately determined temperature is required for each 
ionic species to correctly derive its abundance.
Fluxes of the recombination lines of hydrogen, helium and some heavy 
elements seen in the optical spectra of H {\sc ii} regions are less 
sensitive to electron temperature. However for some scientific problems, 
a very high precision in the element abundances derived from  
recombination lines is required. For example, the helium primordial 
abundance needs to be determined to better than about 1-2\% to yield 
interesting cosmological constraints. 
In those cases, to reach the desired precision, 
the variations of the electron temperature in the 
H {\sc ii} region need to be taken into account for the recombination 
transitions as well.

  There are several known 
methods for the determination of the electron temperature
$T_e$ in H {\sc ii} regions. One of them is based on the 
temperature-sensitive intensity ratios of collisionally excited forbidden
lines of different ions, such as
[O {\sc iii}] $\lambda$(4959+5007)/$\lambda$4363, 
[Ne {\sc iii}] $\lambda$3868/$\lambda$3342,
[Ar {\sc iii}] $\lambda$7135/$\lambda$5192,
[N {\sc ii}] $\lambda$(6548+6584)/$\lambda$5754,
[O {\sc ii}] $\lambda$(7320+7330)/$\lambda$3727 and others \citep{Aller84}.
Most often, it is 
 the nebular to auroral line intensity ratio of doubly ionized 
oxygen [O {\sc iii}] $\lambda$(4959+5007)/$\lambda$4363 that is used. 
Such a method can be applied to H {\sc ii} regions with $T_e$$\ga$10$^4$K
where the auroral emission line [O {\sc iii}] $\lambda$4363 
is sufficiently bright to be detected \citep[e.g. ][]{I05}.

Another way for determining electron temperatures in H {\sc ii} regions
relies on discontinuities in the continuum emission 
of hydrogen and helium free-bound recombination 
processes. Most often the Balmer discontinuity at 3646 \AA\ 
and the Paschen discontinuity at 8207 \AA\ are used. 
\citet{P67} was the first to use the Balmer 
discontinuity, or Balmer jump, to derive electron temperatures 
$T_e$(H$^+$) in several Galactic H {\sc ii} regions, 
using the intensity ratio BJ/$I$(H$\beta$),
where BJ = $I$(3646$^{-}$) -- $I$(3646$^+$) is the value of the Balmer 
jump. Much later, to reduce the interstellar reddening effect
on the determination of $T_e$,
\citet{Liu2000} and \citet{Luo2001} have proposed to use 
the H11 hydrogen emission line instead of the H$\beta$ emission line,
defining the Balmer jump as BJ/H11 = 
[$I_c$($\lambda$3646) -- $I_c$($\lambda$3681)]/$I$(H11 $\lambda$3770). 
  More recently, \citet{Z2004} have suggested a new approach to derive the
electron temperature and density in H {\sc ii} regions
from the observed nebular hydrogen recombination spectrum. The method is based 
on fitting the dereddened observed 
spectrum around the region of the Balmer jump 
with model spectra calculated from recombination theory by varying the 
electron temperature and density. 

While these methods of electron temperature determination may in principle 
be applied to any type of photoionized gaseous region, in practice 
they have been mainly applied to 
planetary nebulae (PNe) \citep[e.g. ][]{Liu2000,Luo2001,R03,Z2004,P04,W05}.
Only a few electron temperature measurements from the Balmer jump 
or Paschen jump techniques 
exist in the literature for H {\sc ii} regions in the Milky Way 
and several nearby galaxies \citep[e.g. ][]{P92,P93,P00,P03,E98,G04,G05,G06}.
For example, \citet{G94} have used the Paschen jump 
to derive the electron temperature from the hydrogen 
recombination spectrum in two H {\sc ii} regions of the low-metallicity
($Z$ $\sim$ $Z_\odot$/10) blue compact dwarf (BCD) galaxy NGC 2366.

 A general finding of the above work for PNe is that 
 the electron temperature $T_e$(H$^+$) derived from the Balmer jump 
is systematically lower than the temperature
$T_e$(O {\sc iii}) derived from the collisionally excited 
[O {\sc iii}] emission lines. Furthermore, \citet{Z2005a,Z2005b} by using     
the He {\sc i} recombination lines and the
He {\sc i} discontinuity at 3421\AA\ to measure electron temperatures, have   
found that $T_e$(He$^+$) in PNe is significantly lower than $T_e$(H$^+$).
As for H {\sc ii} regions, the situation is not so clear.
In some H {\sc ii} regions the electron temperature $T_e$(H$^+$) is
appreciably lower than $T_e$(O {\sc iii}) \citep[e.g., in M17, ][]{P92,P93}
while in some other H {\sc ii} regions, the two
 temperatures are comparable
\citep[e.g., in Orion nebula, ][]{E98}. 

The exact cause of 
these temperature differences is still unclear. 
Following \citet{P67}, they are usually 
attributed to temperature fluctuations in chemically homogeneous gaseous 
nebulae. 
However, no homogeneous photoionization models can give temperature 
fluctuations as large as those observed. If we characterize the spatial 
temperature fluctuations in H {\sc ii} regions by $t^2$,
the mean square temperature variation \citep{P67}, then 
models give $t^2$ in the 0.002 -- 0.025 range, with typical values 
around 0.005, while observations give $t^2$ in the 0.02 -- 0.04 range 
\citep[see the review in ][]{O2003}.     
To reduce the discrepancy between models and observations, 
\citet{V1994} have proposed that chemically homogeneous nebulae contain  
condensations with both temperature and density variations. However,
\citet{Liu2001} have concluded that temperature 
fluctuations alone are unable to explain the large differences between
$T_e$(H$^+$) and $T_e$(O {\sc iii}) in PNe.
Instead, they proposed that PNe contain H-deficient clumps 
with temperatures, densities and chemical compositions different from those 
of the diffuse ionized gas around the clumps.
On the other hand, 
\citet{O2003} have studied small-scale temperature fluctuations in the 
Orion nebula with {\sl HST} images, but did not find statistically
significant variations of the columnar electron temperatures with distance from
the ionizing star, in agreement with photoionization models for chemically
homogeneus gaseous nebulae of constant density. However, small-scale variations
(over a few arcsec or $\sim$ 0.005 pc) were found.
Those authors concluded that none of the physical
mechanisms that have been proposed thus far is able to account 
for the observed temperature differences.

The difference between $T_e$(H$^+$) and $T_e$(O {\sc iii}) may have important 
consequences for the heavy element abundance
determination in H {\sc ii} regions,
as their derived abundances depend sensitively on the adopted temperature.
Indeed, in the majority of such studies
\citep[e.g. ][]{I05}, the heavy element abundances are derived assuming that
the temperature of the H$^+$ zone is equal to the temperature of the O$^{2+}$
zone. If instead, $T_e$(H$^+$) is smaller than $T_e$(O {\sc iii}), then  
the heavy element abundances would be increased.
This issue is especially important for the determination of 
the primordial He abundance from spectra of low-metallicity BCDs. 
In this work \citep{ITL94,ITL97,IT98,OSS97,OS04,IT04a}, it is assumed
that $T_e$(H$^+$) = $T_e$(O {\sc iii}). If indeed  
$T_e$(H$^+$) is less than $T_e$(O {\sc iii}) in the H {\sc ii} regions 
of the low-metallicity BCDs, then the derived He abundances would be lower,
resulting in a lower primordial He abundance \citep{L03}.

In all the above studies, direct measurements of $T_e$(H$^+$) by the 
Balmer or Paschen jump techniques have been carried out only for  
relatively high-metallicity H {\sc ii} regions. The metallicity range 
is generally between 1/3 solar and solar metallicity. With the exception 
of Mrk 71 = NGC 2363 \citep{G94} with a metallicity of 1/10 that of 
the Sun, no direct   
measurements of $T_e$(H$^+$) in very low-metallicity BCDs have been done.
Yet, those very metal-deficient objects which span a metallicity range 
of about 1/50 to 1/10 of the Sun's metallicity are precisely those used in the 
determination of the primordial He abundance. Thus to assess a potential 
source of systematic error in the primordial He abundance measurement 
which may 
compromise the reliability of the cosmological constraints that are  
derived from it, it is crucial 
to investigate whether temperature differences are as important in 
low-metallicity BCDs as they are in high-metallicity H {\sc ii} regions and 
PNe.  

We have thus embarked on an observational program with the MMT
to obtain blue spectra of a large sample of BCDs 
in the 3200 -- 5200\AA\ wavelength range which includes 
the region of the Balmer jump. In addition, we have supplemented 
our data with spectra of bright H {\sc ii} regions from the
Data Release 3 (DR3) of the  Sloan Digital Sky Survey (SDSS) in the
wavelength range 3800 -- 9200\AA, which includes the Paschen jump region.
We wish to directly measure the
electron temperature $T_e$(H$^+$) in these H {\sc ii} regions 
by applying either the Balmer jump technique  
on the MMT spectra or the Paschen jump technique on the SDSS spectra.
We present in this paper the first results of the program. In Section 
\ref{Tee}, we 
describe the observational sample and the method used to determine $T_e$(H$^+$)
in the H {\sc ii} regions. We compare $T_e$(H$^+$) and 
$T_e$(O {\sc iii}) for our sample of H {\sc ii} regions and discuss the 
issue of temperature differences in low-metallicity emission-line galaxies 
in Section 
\ref{results}. We summarize our conclusions in Section \ref{conc}.

\section{Electron temperature determination with the Balmer
and Paschen jumps\label{Tee}} 

\subsection{Observations and data reduction \label{obs}}

The MMT observational sample consists of 22 low-metallicity 
H {\sc ii} regions in 18 BCDs and of one H {\sc ii} region, J1404+5423,
in the spiral galaxy M101. The sample is the same as the one used by  
\citet{TI05} to study high-ionization
emission lines in low-metallicity BCDs. The selection criteria 
of the sample are given in \citet{TI05}, so we will not repeat them 
here in detail. For our present purposes, it is sufficient 
to know that 
the BCDs were chosen to span a large metallicity range, from about 
 1/60 of solar metallicity for SBS 0335--052W  
to about 1/3 of solar metallicity for 
Mrk 35 = Haro 3, so we can study any trend 
of differences between $T_e$(H$^+$) and $T_e$(O {\sc iii}) 
with metallicity. Since the electron 
temperature of an H {\sc ii} region depends on its metallicity, 
this ensures that 
the H {\sc ii} regions in our 
BCD sample span a large range of electron temperatures.
They were also selected to cover a large range of equivalent widths of 
the H$\beta$ emission line, with EW(H$\beta$) varying from 
from 18\AA\ to 397\AA. Since EW(H$\beta$) is a measure of the relative
contribution of the ionized gas emission to the total light, this allows to
study how stellar emission affects the derived electron
temperatures.
The observed BCDs are listed in order of decreasing EW(H$\beta$), i.e. in 
order of increasing contribution of the stellar component to the total light,
 in Table \ref{te}.    
The general characteristics for 
each BCD can be found in Table 1 of \citet{TI05}.
We can also determine electron temperatures from the Paschen jump
at 8207 \AA. To this end,   
we have extracted spectra of 24 bright H {\sc ii} regions
from the DR3 of the SDSS which cover the red wavelength 
region and include the 
Paschen jump, and which constitute 
what will be designated hereafter as the SDSS sample. 
The SDSS sample includes eleven H {\sc ii} regions
in dwarf emission-line galaxies, and 13 H {\sc ii} regions 
in spiral galaxies in a metallicity range from about 1/50 of solar
metallicity for I Zw 18 SE to about 1/2 of solar metallicity for J1402+5416.
The extracted H {\sc ii} regions with their
derived parameters and spectroscopic SDSS ID numbers are listed in order of 
decreasing EW(H$\beta$) in Table \ref{te1}. 
While the metallicity range of the SDSS sample 
is about the same as that of the MMT sample, 
the SDSS sample contains a larger fraction of 
high-metallicity H {\sc ii} regions. 
 
The observations of the MMT sample were carried out with the 6.5-m 
MMT\footnote{The observations
reported here were obtained at the MMT Observatory, a joint facility of the
Smithsonian Institution and the University of Arizona.}
on the nights of 2004 February 19--20 and 2005 February 4, using 
the Blue Channel of the MMT spectrograph. 
We used a
2\arcsec$\times$300\arcsec\ slit and a 800 grooves/mm grating in first order.
The above instrumental set-up gave a spatial scale along the slit of
0\farcs6 pixel$^{-1}$, a scale perpendicular to the slit of 
0.75\AA\ pixel$^{-1}$, a spectral range of 3200 -- 5200\AA\ and a spectral 
resolution of $\sim$ 3\AA\ (FWHM). The seeing was in the range 
1\arcsec--2\arcsec. 
The spectra of the majority of the galaxies have been obtained with the slit
oriented with the parallactic angle. The only exceptions concern 
SBS 0335--052W and 
SBS 0335--052E where the slit was oriented along the major axes of the 
galaxies,
with position angles of --30$^\circ$ and --80$^\circ$ respectively, and
Mrk 71, where the slit was oriented along the position angle of +60$^\circ$, 
so as to go through the three brightest H {\sc ii} regions. In those cases, 
differential refraction may play a role, and  
we analyze its possible effect in Section \ref{results}.
Three Kitt Peak IRS spectroscopic standard stars, G191B2B,
Feige 34 and HZ 44 were observed at the beginning, middle and end of each
night for flux calibration. For the final calibration curve, we adopt the mean 
of three derived response curves, which does not deviate by 
more than 1\% from the response curves obtained from each
standard star individually, over the whole wavelength range.
Spectra of He-Ar comparison arcs were obtained
before and after each observation to calibrate the wavelength scale. The log
of observations is given in Table 2 of \citet{TI05} and 
  details of the data reduction can be found in the same paper.

For the present analysis, 
we have used the same one-dimensional spectra as those
of \citet{TI05}. They have been 
extracted from the two-dimensional spectra using 
a 6\arcsec $\times$ 2\arcsec\ extraction aperture so as to cover the brightest 
parts of the H {\sc ii} regions.
In the case of
five BCDs, Mrk 71, I Zw 18, Mrk 35, Mrk 59 and Mrk 178, spectra of 2-3 
different H {\sc ii} regions in the same galaxy have been 
extracted.
In total, \citet{TI05} extracted 27 one-dimensional spectra.
However, since the spectra of I Zw 18C, Mrk 35 \#2, Mrk 178 \#2 and Mrk 178
\#3 are too noisy for our present purposes \citep[see their
spectra in Fig. 2 of ][]{TI05}, we do not include them here.

  The spectra for all 23 H {\sc ii} regions in the MMT  
sample are displayed in Figure~\ref{sp_2}. We have also shown a zoom
 of the Balmer jump region for four selected H {\sc ii} regions 
in Fig. \ref{sp_3}. 
We show the spectra for the SDSS H {\sc ii} regions in Fig. \ref{spp_1}. 
Since we are concerned with the Paschen jump region, 
we have displayed only the 
red part of each spectrum. 
All spectra have been reduced to zero redshift.
The electron temperature $T_e$(O {\sc iii}), ionic and total element
abundances of the H {\sc ii} regions in both the MMT and SDSS samples have 
been derived according to the prescriptions of \citet{I05}. They are given in 
\citet{TI05} for the H {\sc ii} regions in the MMT sample 
and in \citet{I05} for the SDSS sample. 
Relevant parameters such as the 
equivalent width EW(H$\beta$), the oxygen abundance 12 + log O/H,
and the extinction coefficient $C$(H$\beta$)
are given in Table \ref{te} for the MMT objects and in Table \ref{te1} 
for the SDSS objects.

\subsection{Monte Carlo calculations \label{method}}
  The method consists of deriving the electron temperature $T_e$(H$^+$) 
of H$^+$ zones
by fitting a series of model spectral energy distributions (SEDs) to the 
observed ones and finding the best fits. The fit for each MMT spectrum 
is performed over the whole observed spectral range shortward of  
$\lambda$5200\AA, which includes the Balmer jump region ($\lambda$3646\AA).
As for the SDSS sample, each fit is performed over a spectral range of 
3900 -- 9200\AA, which includes the Paschen jump region ($\lambda$8207\AA).
Besides the electron temperature $T_e$(H$^+$) which controls the 
magnitudes of the Balmer and Paschen jumps, the shape of the 
SED depends on several other parameters.

As each SED is the sum of both stellar and ionized gas
emission, its shape depends on the relative contribution of these 
two components. In BCDs, the contribution of the  
ionized gas emission can be very large. However,  
the equivalent widths of the hydrogen 
emission lines never reach the theoretical values for
pure gaseous emission, implying a non-zero contribution of stellar
emission in all objects. 
The contribution of gaseous emission relative to stellar 
emission can be parametrized by 
the equivalent width of the H$\beta$ emission line EW(H$\beta$).
Given a temperature, the ratio of the gaseous 
emission to the total emission 
is equal to the ratio of the observed EW(H$\beta$) to the 
equivalent width of H$\beta$ expected for pure gaseous emission.
The shape of the spectrum depends also on reddening.
The extinction coefficient for the ionized gas $C$(H$\beta$)$_{gas}$ 
can be obtained from 
the observed hydrogen Balmer decrement. However, there is no direct 
way to derive the 
extinction coefficient $C$(H$\beta$)$_{stars}$ for the 
stellar emission which can be different from
the ionized gas extinction coefficient. 
Finally, the SED depends on the star formation history of the galaxy.


We have carried out a series of Monte Carlo simulations 
to reproduce the SED 
in each H {\sc ii} region of our sample. 
   To calculate the contribution of the stellar emission to the SEDs,
we have adopted the grid of the Padua stellar evolution
models by 
\citet{Gi00}\footnote{http://pleiadi.pd.astro.it.} 
with heavy
element mass fractions $Z$ = 0.0001, 0.0004, 0.001, 0.004, 0.008.
To check the sensitivity of our calculations to the particular stellar tracks 
adopted, we have also considered Geneva tracks \citep{LS01}. 
With these tracks, we obtained results similar to those gotten 
with the Padua ones. Nevertheless, the Geneva models are less appropriate for 
our goals as they do not include stellar models older than 1 Gyr. 
Therefore we do not include the results obtained  
with the Geneva models in this paper.
To reproduce the SED of the stellar component with any star formation 
history, we have calculated with the
package PEGASE.2 \citep{FR97} a grid of  
instantaneous burst SEDs in a wide range of ages, from  0.5 Myr to 15 Gyr.
 We have adopted a stellar initial mass function with a Salpeter 
slope, an
upper mass limit of 100 $M_\odot$ and a lower mass limit of 0.1 $M_\odot$.
Then the SED with any star formation history can be obtained by integrating
the instantaneous bursts SEDs over time with a specified 
time-varying star formation rate.

More than $\sim$90\% of the emission of an H {\sc ii} region 
is due to the ongoing burst of star formation. Therefore, the shape of its 
spectrum is not very sensitive to variations in the star formation history of 
the older stellar populations. Thus,  
as a first approximation, we have approximated the  
star formation history in each BCD   
by two short bursts with different strengths: a recent 
burst with age $t$(young) $<$ 10 Myr which accounts for the 
young stellar population and a prior burst 
with age $t$(old) varying 
between 10 Myr and 15 Gyr and responsible for the older stars. 
   The contribution of each burst
to the SED is defined by the ratio of the masses of stellar populations
formed respectively 
in the old and young bursts, $b$ = $M$(old)/$M$(young), which we vary 
between 0.01 and 100. 
To check the sensitivity of our calculations to the adopted 
star formation history, we have also considered the case where 
star formation is continuous for the old stellar population. 

The contribution of the gaseous emission was scaled 
to the stellar emission using the ratio of the observed equivalent width of 
the H$\beta$ emission line to the equivalent width of H$\beta$ expected 
for pure gaseous emission. 
The continuous gaseous emission is taken from \citet{Aller84} and 
includes hydrogen and helium free-bound, free-free and 
two-photon emission. In our models it is always calculated with the electron
temperature $T_e$(H$^+$) of the H$^+$ zone and with the chemical composition 
derived from the H {\sc ii} region spectrum.
The observed emission lines corrected for reddening
and scaled using the absolute flux of the H$\beta$ emission line were added to 
the calculated gaseous continuum. 
  The fraction of the gaseous continuum to the total light depends on the 
adopted electron temperature $T_e$(H$^+$) in the H$^+$ zone, since EW(H$\beta$)
for pure gaseous emission decreases with increasing $T_e$(H$^+$).
Given that $T_e$(H$^+$) is not necessarily
equal to $T_e$(O {\sc iii}), we chose to vary it in the range 
(0.7 -- 1.3)$\times$$T_e$(O {\sc iii}). 
  Strong emission lines in BCDs (see Fig.~\ref{sp_1}) are
usually measured with good precision and so the equivalent 
width of the H$\beta$ emission line and the extinction coefficient
for the ionized gas are accurate to $\pm$5\% or $\pm$20\%, respectively.
Thus we vary EW(H$\beta$) between 0.95 and 1.05 times its nominal value.
As for the extinction, we assume that
the extinction $C$(H$\beta$)$_{stars}$ for the stellar light is the same as 
 $C$(H$\beta$)$_{gas}$ for the ionized gas,
and we vary both in the range (0.8 -- 1.2)$\times$$C$(H$\beta$)$_{gas}$.
  We run a large number (either 10$^5$ or 10$^6$) of Monte Carlo models 
varying randomly $t$(young), $t$(old), $b$ and $T_e$(H$^+$)
in a large range, and
EW(H$\beta$), $C$(H$\beta$)$_{gas}$ and $C$(H$\beta$)$_{stars}$ in a
relatively smaller range because the latter quantities are more 
directly constrained by observations. 
Because of the somewhat large number of parameters, the
determination of $T_e$(H$^+$) in any particular H {\sc ii} region may not 
be very certain. In addition to model assumptions, the result may also
be affected by observational and data reduction uncertainties. 
Therefore it is 
important to adopt a statistical approach and 
use a large sample of H {\sc ii} regions covering a wide range
of metallicities to make more definite conclusions on the temperature  
differences between the O$^{2+}$ and H$^+$ zones.

\subsection{Goodness of fit}

To judge the goodness of each model's fit to the observed SED, 
we have computed a mean deviation $\sigma$ between the observed and modeled
spectra for each Monte Carlo realization. To obtain $\sigma$, 
we calculate the root mean square deviations $\sigma_j$ for  
each of five continuum regions chosen to cover the whole wavelength range and 
selected to be devoid of line emission in each spectrum. Two of the selected 
regions are chosen to be on either side of the Balmer jump 
and another two on either side of the Paschen jump.
They are chosen to be as close as possible to the locations of the two jumps.
The five chosen continuum regions are shown by thick horizontal lines
in representative 
MMT and SDSS spectra in Fig. \ref{spcgcg}, along with the best model fit to 
the whole spectrum. In the spectra of other objects, the precise location of 
the continuum regions 
may vary slightly, depending on the quality of data, the presence 
of residuals from night sky subtraction, etc.
The quantity $\sigma_j$ is calculated as $\sigma_j$  =
$ \sqrt{\sum_{i=1}^{N_j}{(f^i_{obs}-f^i_{mod})^2} / N_j}$, where $N_j$ is 
the number of points in each particular spectral interval. Then 
$\sigma$ = $\sum{\sigma_j}/5$. 
 We did not attempt to fit stellar
absorption features such as the hydrogen absorption lines 
 because of their weakness, especially in the lower 
signal-to-noise ratio SDSS spectra.
Additionally, the broad wings of the 
strong hydrogen emission lines significantly
alter the shape of the absorption lines, especially in H {\sc ii} regions
with high EW(H$\beta$), such as Mrk 71, II Zw 40, SBS 0335--052E and others
(see Fig. \ref{sp_1}).
 However, in
galaxies with relatively low EW(H$\beta$), the hydrogen absorption features
are reproduced quite well (see for example the case of I Zw 18 SE in 
Fig. \ref{sp_3}).

To gauge the goodness of the fit, we have used the 
$\sigma$ statistic instead of the usual 
$\chi^2$ test because in the latter statistic, the weight of each region in 
the overall SED fit depends on the spectrum noise in it. 
Since the noise is largest in the bluest part of the MMT spectrum
which includes the Balmer jump, the weight of this region in the 
fit is always small. 
Therefore, the Balmer jump region which plays a crucial role for the  
determination of $T_e$(H$^+$) is not well fitted by the model SED 
if the $\chi^2$ statistic is used. For uniformity, the 
same $\sigma$ statistic has been adopted for the SDSS spectra.

In Figure \ref{dmetal} we show the parameter space covered by our 
Monte Carlo simulations for two BCDs observed with the MMT 
which are at the extreme ends of  
the metallicity range of our sample: 
Mrk 35 which is the highest-metallicity BCD in the MMT sample
(upper panel) and SBS 0335--052E which is  
one of the lowest-metallicity BCDs in it
(lower panel). A detailed presentation of the model results for  
the whole MMT sample will be given in a subsequent paper. There, 
the star formation 
history and the stellar populations of each object will be discussed. 
Calculations have been done for heavy element mass 
fractions of the stellar populations closest to the heavy element
mass fraction of the ionized gas: $Z_\odot$/2.5 (Fig. \ref{dmetal}a,b)
for Mrk 35 and $Z_\odot$/50 (Fig. \ref{dmetal}c,d) for 
SBS 0335--052E. In Fig. \ref{dmetal}a and \ref{dmetal}c, we show only those 
Monte Carlo realizations that yield a EW(H$\beta$) within
$\pm$ 5\% of the observed one, i.e. 2650 out of 10$^5$ 
realizations for Mrk 35 and 13972 out of 10$^6$ realizations 
for SBS 0335--052E. In those figures, $t_e$(H$^+$) = $T_e$(H$^+$)/
10000.
 It is seen that the age
$t$(old) of the old stellar population, the age $t$(young) of the young
stellar population, the ratio $b$ of their masses and the  
electron temperature $T_e$(H$^+$) in the H$^+$ zone vary in a wide range.
However, if only the 10 best solutions with the lowest $\sigma$s are 
considered, then the range of the electron temperature 
in the H$^+$ zone (Fig.~\ref{dmetal}b and \ref{dmetal}d), 
is considerably narrowed. The adopted temperature is then taken to be the 
mean of these 10 best temperatures, weighted by $\sigma$.

Some other parameters which define the 10 best solutions lie also in a
relatively narrow range and their values shown in Fig.~\ref{dmetal}b and 
\ref{dmetal}d for Mrk 35 and SBS 0335--052E 
are representative for the whole sample. In particular,
$t$(young) lies in a very narrow range because it depends very sensitively on  
EW(H$\beta$) and is strongly constrained by it.  
The ratio $b$ of the masses of old to young stellar populations is typically 
$\la$ 10, corresponding to a luminosity ratio of $\la$ 10\%.
As for $t$(old), its range in 
the 10 best solutions is larger, but is typically between 100 Myr and several
Gyr for the whole H {\sc ii} region sample.

We have checked that the derived $t_e$(H$^+$) does not 
depend sensitively on the exact number of best solutions we choose to 
average.  
 Fig.~\ref{t_10_20} shows 
the mean temperature $t_e$(H$^+$) derived from the 10 
best fits vs. the one derived from the 
20 best fits for all H {\sc ii} regions in our sample. 
No systematic deviation from the perfect equality line (dashed line) 
is seen. We will thus adopt $t_e$(H$^+$) as the weighted mean of the 10 best 
solutions.

\section{Results \label{results}}

\subsection{Accuracy of the derived temperatures}

Figs.~\ref{sp_2} -- \ref{spp_1} show by a thick solid line 
the best fit SED (the one with the smallest $\sigma$)
superimposed on the redshift- and 
extinction-corrected observed spectrum of every H {\sc ii} region 
in the MMT and SDSS samples. 
In every case, the model SED fits well the observed one over 
the whole spectral range, including the Balmer or the 
Paschen jump region.  
The separate contributions 
from the stellar and ionized gas components are shown
by thin solid lines.
 The H {\sc ii} regions are arranged in order of decreasing of H$\beta$ 
equivalent width, i.e. in order of decreasing contribution of 
gaseous emission relative to stellar emission. The $t_e$(H$^+$) that are 
derived from the best fits are given in Table \ref{te} for the MMT sample,
and in Table \ref{te1} for the SDSS sample 
 in the column labeled ``inst +inst '' (to denote instantaneous bursts for 
both the young and old stellar populations). We 
examine next several factors which may affect the electron 
temperatures so derived.  

\subsubsection{Differential refraction}

We have noted in Sect. \ref{Tee} that the majority of the H {\sc ii}
regions from the MMT sample were observed at parallactic angle. Therefore,
differential refraction is not important for those objects. However that was 
not the case for 3 objects, SBS 0335--052W, SBS 0335--052E and Mrk 71. 
To check for  
differential refraction effects in those cases, we have 
modeled every of the seven individual 
spectra of SBS 0335--052E obtained at different airmasses. The weighted 
mean temperature $t_e$(H$^+$) of the 10 
best SED fits for each observation is plotted as a function of airmass 
in Fig. \ref{atmref}. There is no evident trend.
The absence of dependence of $t_e$ on airmass can probably be explained by the 
fact that the reddening correction applied to the whole spectrum 
is based on the observed fluxes of the hydrogen 
Balmer lines which are themselves subject to differential refraction effects.
That procedure thus automatically removes a large part of such effects  
 in the observed SEDs. Therefore we 
conclude that differential refraction does not significantly affect  
the electron temperature determination in the three H {\sc ii} regions in 
our MMT sample that are subject to it.  The same conclusion holds for SDSS 
spectra as well. These were obtained with an airmass not exceeding 1.3.

\subsubsection{Continuous star formation}

All of our above model fitting was done assuming that the stellar 
populations
contributing to the light of the H {\sc ii} region were formed in two
short bursts of star formation, a recent one for the young stellar 
population, and a prior one for the old stellar population.
  While the approximation of an 
instantaneous burst is a good one for the young ionizing stellar 
population \citep[e.g. ][]{Papad98,Fricke2001,Noes00}, it is probably less so 
for the old stellar population.
The observed SED of the old stellar population in the 
underlying extended component of a galaxy is more realistically modeled 
by a continuous star formation, as shown e.g. by 
\citet{Guseva2001,Guseva2003a,Guseva2003b,Guseva2003c}. 
We wish to investigate how our derived temperatures depend on 
the adopted star formation history in a galaxy. To this end, we 
have also carried out Monte Carlo simulations for all objects 
in the MMT and SDSS samples in which 
the old stellar population is assumed to be formed not in a short burst, but 
continuously with a constant star formation rate 
starting at an initial time $t_i$ and
stopping at a final time $t_f$. This is similar to the approach adopted by  
\citet{Guseva2001,Guseva2003a,Guseva2003b,Guseva2003c}. 
Since the exact star formation history for the old stellar population is
not known, we choose $t_i$ and $t_f$ to be random numbers 
between 10 Myr and 15 Gyr with the condition 
that $t_f$ $\leq$ $t_i$. 
The best solutions are derived in the same way as for 
the two burst model simulations.
The results of our calculations are shown
in Tables \ref{te} and \ref{te1} in the column labeled 
``inst + cont'' (to denote that the young stellar population is 
formed in an instantaneous burst and the old one in a continuous mode). 
 Examination of Fig. \ref{t_inst_cont} where $t_e$(inst + inst) is plotted 
against $t_e$(inst + cont) reveals no significant 
difference between the two temperatures. 
The relative insensitivity of the derived temperatures $t_e$(H$^+$) 
to the particular star formation history of the old 
stellar population is due to the fact that the contribution of this population 
to the light of the bright H {\sc ii} region
is always small, not exceeding a few percent in the optical range.  
We conclude that the two-burst model is a good approximation to the type of 
objects considered here.
Therefore, in the following, we will consider only  
results obtained with the two-burst
models. They require much less computational time than those 
with continuous star formation for the old stellar population.

\subsubsection{Dependence on the contribution of the gaseous emission and on 
the oxygen abundance} 

We show respectively in Figs.~\ref{t_e} and \ref{t_ep} by filled circles the 
temperature $t_e$(H$^+$) derived from each of the 10 best SED fits for 
the H {\sc ii} regions in the MMT and SDSS samples. 
The $\sigma$ weighted mean of these temperatures
is indicated by a straight vertical line in each panel.
This mean temperature $t_e$(H$^+$) is listed in 
Tables ~\ref{te} and \ref{te1} for every 
H {\sc ii} region, along with 
the temperature $t_e$(O {\sc iii}) = 10$^{-4}$ $T_e$(O {\sc iii})  
derived from the [O {\sc iii}] $\lambda$(4959+5007)/$\lambda$4363
nebular to auroral intensity line ratio.
It can be seen from Figs.~\ref{t_e} and \ref{t_ep} that in some cases, 
Mrk 71A, J0519+0007, HS 0837+4717, HS 0822+3542, 
SBS 0335--052E and Mrk 35 in Fig. \ref{t_e}, and Mrk 1236,
J1402+5416, J1402+5414,
J1304--0333, J1228+4407, J1404+5425 and J1403+5421 in Fig. \ref{t_ep}, 
the temperature dispersion is small, corresponding to a relatively
deep minimum 
of $\sigma$ values in the parameter space and a well-defined $t_e$(H$^+$).
In other cases however, the dispersion of the temperatures 
is larger, corresponding to a broader minimum of $\sigma$ values 
and a more uncertain $t_e$(H$^+$).

We wish to examine next 
which factors may affect the dispersion of $t_e$(H$^+$) and hence limit
the precision of its determination. A good measure of that dispersion is the 
difference between the maximum electron temperature $t_{max}$ and
the minimum electron temperature $t_{min}$ of the 10 best solutions.
In Fig.~\ref{delta_t} we show the dependence of 
the difference $t_{max}$ -- $t_{min}$ on the equivalent width of 
H$\beta$ (Fig.~\ref{delta_t}a), the oxygen abundance 12 + log(O/H) 
(Fig.~\ref{delta_t}b) and the interstellar extinction $C$(H$\beta$)
(Fig.~\ref{delta_t}c) for the whole sample. 
 We find an anti-correlation between $t_{max}$ -- $t_{min}$ and 
EW(H$\beta$) with however a large dispersion (Fig.~\ref{delta_t}a). 
  This implies that, in general,
 the uncertainty in the $t_e$(H$^+$) determination is
smaller for galaxies with a larger contribution of gas emission since
the effect of the stellar SED uncertainties is less in these galaxies. 
 A small anti-correlation between $t_{max}$ -- $t_{min}$
and the oxygen abundance (Fig.~\ref{delta_t}b) may be present. 
This slight trend, if it exists,
 may be explained by more reliable measurements of the gaseous
Balmer or Paschen jumps at higher metallicities.
High metallicities lead to lower temperatures of the 
ionized gas and hence to larger gaseous Balmer and Paschen 
jumps which can be modeled more accurately.
  On the other hand, there is no 
clear dependence of $t_{max}$ -- $t_{min}$ on
interstellar extinction (Fig.~\ref{delta_t}c).

\subsection{Comparison of $t_e$(H$^+$) with $t_e$(O {\sc iii})}

We now compare $t_e$(H$^+$) of the H$^+$ zone derived from the Balmer or 
Paschen jumps and $t_e$(O {\sc iii}) of the O$^{2+}$ 
zone derived from the nebular to auroral line intensity ratio of doubly 
ionized oxygen [O {\sc iii}] $\lambda$(4959+5007)/$\lambda$4363. 
In Fig.~\ref{tP0}a and \ref{tP0}b, we plot  
$t_e$(H$^+$) against $t_e$(O {\sc iii}) for all the objects in 
our sample respectively for the two-burst and burst+continuous models.
The MMT objects are denoted by filled circles and the SDSS objects by 
open circles. 
In Fig.~\ref{tP0}c and \ref{tP0}d, the same plots are shown 
but only for objects with
relatively deep minima, i.e. those with 
$\sigma$[$t_e$(H$^+$)]/$t_e$(H$^+$) $<$ 10\%.
We have assumed  
the metallicities of the stars and gas to be the same, an assumption which we 
relax later. 
Because of the discreteness in $Z$ of the grid of 
stellar evolution models, the actual metallicity of the stars
may be slightly greater or smaller than the measured ionized gas 
metallicity. In any case, it is always 
the one in the grid which is closest to the metallicity of the ionized gas. 
As before, $t_e$(H$^+$) for each galaxy is the mean of the temperatures 
derived from the 10 best Monte Carlo realizations weighted by $\sigma$.
The error bars for $t_e$(H$^+$) show the dispersion from the mean.
 The solid line denotes equal temperatures while
the dashed lines illustrate 
$\pm$5\% deviations from the line of equal temperatures.
Examination of Fig. \ref{tP0} shows there are no systematic differences  
between the electron temperature $t_e$(O {\sc iii}) deduced from the 
collisionally excited [O {\sc iii}]  nebular-to-auroral forbidden line ratio 
and the temperature $t_e$(H$^+$) derived from 
the H {\sc i} recombination spectrum.
There are also no systematic differences between
the two-burst and the burst+continuous models, either for the
whole sample (Figs. \ref{tP0}a and \ref{tP0}b) or for H {\sc ii} regions
with deep minima of $t_e$(H$^+$) (Figs. \ref{tP0}c and \ref{tP0}d).

We have also considered the case where  
the stellar metallicity is less than the ionized gas metallicity.
\citet{Schaerer2000} and 
\citet{Guseva1998} have suggested that there is  
some hint of a lower metallicity of the stellar population
as compared to the metallicity of the ionized gas. 
To carry out the modeling, we adopt for the metallicity of the stellar 
population 
the value of $Z$ in the grid of the stellar evolutionary
models which is closest to but lower than the stellar 
abundance adopted in the $Z_{stars}$ = $Z_{gas}$ case. 
The results we obtain for the $Z_{stars}$ $<$ $Z_{gas}$ case 
are very similar to the ones for the $Z_{stars}$ = $Z_{gas}$ case: there are 
no systematic differences between $t_e$(H$^+$) and $t_e$(O {\sc iii}).

For comparison, we have also plotted in Fig.~\ref{tP0}a data from the 
literature for H {\sc ii} regions that possess both $t_e$(H$^+$) 
and $t_e$(O {\sc iii}) determinations. We have shown 
by stars in order of increasing $t_e$(H$^+$) the 
H {\sc ii} regions  M16, M20 and NGC 3603 \citep{G06}, 
the H {\sc ii} region  M17 \citep[two data points, ][]{P92}, 
the H {\sc ii} regions the Orion nebula
\citep[also two data points, ][]{E98}, the H {\sc ii} regions 
NGC 3576 \citep{G04}, S 311 \citep{G05},
30 Dor \citep{P03}, NGC 346 \citep{P00} and the two H {\sc ii} regions 
B and A in the BCD Mrk 71 \citep{G94}. The electron temperatures 
$t_e$(H$^+$) derived for regions A and B in Mrk 71 
by \citep{G94} using the Paschen jump are some 6000K and 4000K
smaller than our values derived from the Balmer jump, leading them 
to conclude than $t_e$(H$^+$) is smaller than $t_e$(O {\sc iii}). We find 
in the contrary that $t_e$(H$^+$) is larger than $t_e$(O {\sc iii}) for
Mrk 71A and equal to $t_e$(O {\sc iii}) for Mrk 71B. 
For comparison with the H {\sc ii} region data, 
we have also plotted in Fig.~\ref{tP0}b the data of \citet{Liu2004}, 
\citet{Z2004} and \citet{W05} for 54 Galactic PNe (dots).
It is seen that the  $t_e$(H$^+$) -- $t_e$(O {\sc iii}) relationship for PNe  
curves down for objects with $t_e$(O {\sc iii}) $\la$ 1.1, with 
$t_e$(H$^+$) being systematically lower than $t_e$(O {\sc iii}), while the 
H {\sc ii} region data do not appear to do so except for the 
lowest-temperature ones.
It is not clear why the differences between $t_e$(H$^+$) and 
$t_e$(O {\sc iii}) in low-temperature PNe are substantial, while they 
appear to be less in  H {\sc ii} regions in the same temperature range.
These differences 
may be caused by chemical abundance inhomogeneities in PNe as discussed by 
\citet{Liu2001}. If so, these abundance inhomogeneities are absent in 
the H {\sc ii} regions in both our MMT and SDSS samples.

There is thus no apparent offset of the $T_e$(H$^+$)
vs. $T_e$(O {\sc iii}) relation from the $T_e$(H$^+$) = $T_e$(O {\sc iii})
line in Fig. \ref{tP0}.
For our sample of 47 H {\sc ii} regions, the
temperature of the O$^{2+}$ zone is on average equal, within the errors, 
to the temperature of the H$^+$ zone.
We emphasize that this is a statistical 
conclusion. Due to the large observational uncertainties for each 
individual object and those introduced 
by the assumed star formation history scenario for 
the modeling of each SED, 
only a statistical study of a large sample of H {\sc ii} regions 
can allow us to draw such a conclusion.
For hot low-metallicity H {\sc ii} regions, we cannot exclude the 
presence of small differences of
$\sim$ 3\% -- 5\% between the temperatures of the O$^{2+}$ zone
and H$^+$ zones, as predicted by \citet{P02}, since the scatter of the data 
in Fig. \ref{tP0} is of that order. In lower temperature H {\sc ii} regions 
however, we do not observe the larger differences of $\sim$ 10\% 
predicted by \citet{P02}. They remain at the level of 
$\sim$ 3\% -- 5\%.

It thus appears that the well-known problem of temperature fluctuations
raised by \citet{P67} affects  
mainly cool high-metallicity PNe, and to a lesser degree   
H {\sc ii} regions. As the determination 
of the primordial He abundance is based mainly on hot low-metallicity
H {\sc ii} regions 
\citep[e.g. ][]{IT04a}, differences between $T_e$(H$^+$)
and $T_e$(O {\sc iii})
of the order of 3\% -- 5\% should not be a main 
concern for the derived value of the primordial He abundance. 
With $T_e$(H$^+$) varying in the range 0.95 -- 1 $T_e$(O {\sc iii}),
Izotov, Thuan \& Stasinska (in preparation, 2006) have shown that the value 
of the primordial helium abundance would be smaller by $\leq$ 1\% than the
value derived assuming $T_e$(H$^+$) = $T_e$(O {\sc iii}).

 \section{Conclusions \label{conc}}

We have used the Balmer and Paschen jumps to 
determine the temperatures of the H$^+$ zones for a total sample of 
47  H {\sc ii} regions. The Balmer jump was used on  
MMT spectrophotometric data of a sample that includes   
22 H {\sc ii} regions in 18 blue compact dwarf (BCD) galaxies and one
H {\sc ii} region in the spiral galaxy M101. 
The Paschen jump was used on spectra of 24 H {\sc ii} regions selected from
the Data Release 3 of the Sloan Digital Sky Survey.
We have employed a Monte Carlo technique
to fit the spectral energy distributions of the galaxies. 
The main conclusions of the study can be summarized as follows:

1. The electron temperature determination of the H$^+$ zone
by fitting SEDs is more reliable for the galaxies with a larger contribution of
the gas emission, i.e. for those with larger EW(H$\beta$).

2. We find that, for our sample of H {\sc ii} regions, 
the temperature of the O$^{2+}$ zone 
determined from the nebular to auroral line 
intensity ratio of doubly ionized 
oxygen [O {\sc iii}] $\lambda$(4959+5007)/$\lambda$4363 is, 
within the scatter of the data, 
equal to the temperature of the H$^+$ zone determined from 
fitting the Balmer jump and the SED. We cannot however rule out 
small temperature differences of the order 3--5\%.

\acknowledgements
The MMT time was available thanks to a grant from the Frank Levinson Fund 
of the Peninsula Community Foundation to the Astronomy Department of the 
University of Virginia.
Y.I.I. and T.X.T. have been partially 
supported by NSF grant AST-02-05785.
The research described in this paper was made possible in part 
by Award No. UP1-2551-KV-03 of the US Civilian Research \& 
Development Foundation for the Independent States of the Former Soviet Union (CRDF). N.G.G. and Y.I.I. are grateful for the hospitality of the Astronomy
Department of the University of Virginia. T.X.T. thanks the hospitality of the
Main Astronomical Observatory in Kiev, of the Institut d'Astrophysique in 
Paris and of the Service d'Astrophysique at Saclay during his sabbatical 
leave. He is grateful for a Sesquicentennial Fellowship from the 
University of Virginia. 
All the authors acknowledge the work of the Sloan Digital Sky
Survey (SDSS) team.
Funding for the SDSS has been provided by the
Alfred P. Sloan Foundation, the Participating Institutions, the National
Aeronautics and Space Administration, the National Science Foundation, the
U.S. Department of Energy, the Japanese Monbukagakusho, and the Max Planck
Society. The SDSS Web site is http://www.sdss.org/.
     The SDSS is managed by the Astrophysical Research Consortium (ARC) for
the Participating Institutions. The Participating Institutions are The
University of Chicago, Fermilab, the Institute for Advanced Study, the Japan
Participation Group, The Johns Hopkins University, the Korean Scientist Group,
Los Alamos National Laboratory, the Max-Planck-Institute for Astronomy (MPIA),
the Max-Planck-Institute for Astrophysics (MPA), New Mexico State University,
University of Pittsburgh, University of Portsmouth, Princeton University, the
United States Naval Observatory, and the University of Washington.


%

\clearpage


\begin{deluxetable}{llccccc}
\tablenum{1}
\tablecolumns{7}
\tablewidth{0pt}
\tablecaption{Comparison of $t_e$(O {\sc iii})\tablenotemark{a} 
and $t_e$(H$^+$)\tablenotemark{b} for the sample galaxies observed with
the MMT.
\label{te}}
\tablehead{
\colhead{Galaxy} &\colhead{EW(H$\beta$)\tablenotemark{c}} &\colhead{$t_e$(O {\sc iii})} 
& \multicolumn{2}{c}{$t_e$(H$^+$)} &\colhead{12 + log(O/H)} 
& \colhead{$C$(H$\beta$)$_{gas}$\tablenotemark{c}} \\ \cline{4-5}
 & & & \colhead{inst+inst}& \colhead{inst+cont}& & 
}
\startdata
Mrk 71A       & 397 & 1.57$\pm$0.01 & 1.87$\pm$0.01 & 1.85$\pm$0.03& 7.87 &0.16  \\
Mrk 94        & 296 & 1.27$\pm$0.01 & 1.33$\pm$0.12 & 1.34$\pm$0.14& 8.11 &0.10  \\
II Zw 40      & 258 & 1.30$\pm$0.01 & 1.48$\pm$0.21 & 1.39$\pm$0.16& 8.13 &0.96  \\
J 0519+0007   & 235 & 1.92$\pm$0.02 & 1.96$\pm$0.04 & 1.98$\pm$0.05& 7.47 &0.13  \\
HS 0837+4717  & 235 & 1.91$\pm$0.02 & 1.95$\pm$0.06 & 2.03$\pm$0.22& 7.60 &0.16  \\
HS 0822+3542  & 232 & 1.82$\pm$0.02 & 1.94$\pm$0.05 & 1.91$\pm$0.09& 7.46 &0.00  \\
J 1404+5423   & 228 & 1.34$\pm$0.01 & 1.41$\pm$0.16 & 1.47$\pm$0.13& 8.06 &0.08  \\
Mrk 209       & 212 & 1.56$\pm$0.01 & 1.65$\pm$0.29 & 1.58$\pm$0.17& 7.82 &0.09  \\
SBS 0940+544  & 207 & 1.90$\pm$0.03 & 1.85$\pm$0.24 & 1.89$\pm$0.30& 7.48 &0.20  \\
SBS 0335-052E & 190 & 2.00$\pm$0.02 & 1.97$\pm$0.05 & 1.80$\pm$0.11& 7.31 &0.14  \\
Mrk 71B       & 181 & 1.47$\pm$0.01 & 1.46$\pm$0.16 & 1.48$\pm$0.17& 7.91 &0.20  \\
SBS 1152+579  & 169 & 1.54$\pm$0.02 & 1.67$\pm$0.25 & 1.56$\pm$0.08& 7.89 &0.18  \\
Mrk 35        & 145 & 1.01$\pm$0.01 & 1.09$\pm$0.08 & 1.08$\pm$0.08& 8.34 &0.14  \\
Mrk 71C       & 127 & 1.39$\pm$0.02 & 1.39$\pm$0.16 & 1.47$\pm$0.17& 7.91 &0.19  \\
Mrk 59(\#1)   & 125 & 1.35$\pm$0.01 & 1.37$\pm$0.26 & 1.37$\pm$0.12& 8.01 &0.02  \\
I Zw 18SE     & 124 & 1.81$\pm$0.03 & 1.82$\pm$0.28 & 1.76$\pm$0.22& 7.22 &0.24  \\
SBS 0926+606A & 116 & 1.31$\pm$0.01 & 1.53$\pm$0.12 & 1.32$\pm$0.15& 8.01 &0.08  \\
SBS 0911+472  & 98  & 1.28$\pm$0.01 & 1.31$\pm$0.16 & 1.34$\pm$0.15& 8.09 &0.20  \\
SBS 1030+583  &  83 & 1.54$\pm$0.02 & 1.35$\pm$0.20 & 1.58$\pm$0.13& 7.81 &0.04  \\
I Zw 18NW     &  81 & 2.00$\pm$0.03 & 1.99$\pm$0.38 & 1.94$\pm$0.20& 7.17 &0.06  \\
SBS 0335-052W &  80 & 1.98$\pm$0.13 & 1.74$\pm$0.21 & 1.90$\pm$0.18& 7.11 &0.12  \\
Mrk 178(\#1)  &  27 & 1.47$\pm$0.03 & 1.44$\pm$0.19 & 1.51$\pm$0.17& 7.93 &0.04  \\
Mrk 59(\#2)   &  18 & 1.38$\pm$0.03 & 1.41$\pm$0.19 & 1.33$\pm$0.13& 8.03 &0.00  \\
\enddata
\tablenotetext{a}{$t_e$(O {\sc iii}) = 
$T_e$(O {\sc iii})/10000 K, derived from the [O {\sc iii}] $\lambda$(4959+5007)/$\lambda$4363 emission line
ratio.}
\tablenotetext{b}{$t_e$(H$^+$) = $T_e$(H$^+$)/10000, derived as the 
weighted mean temperature from the 10 best fits of the SED.}
\tablenotetext{c}{Equivalent widths of the H$\beta$ emission line, 
oxygen abundances and extinction coefficients derived from the 
observed Balmer decrement are taken from \citet{TI05}.}
\end{deluxetable}

\clearpage


\begin{deluxetable}{llcccccc}
\rotate
\small
\tablenum{2}
\tablecolumns{8}
\tablewidth{0pt}
\tablecaption{Comparison of $t_e$(O {\sc iii})\tablenotemark{a} 
and $t_e$(H$^+$)\tablenotemark{b} for the SDSS galaxies.
\label{te1}}
\tablehead{
\colhead{Galaxy} &\colhead{EW(H$\beta$)\tablenotemark{c}} &\colhead{$t_e$(O {\sc iii})} 
& \multicolumn{2}{c}{$t_e$(H$^+$)} &\colhead{12 + log(O/H)} 
& \colhead{$C$(H$\beta$)$_{gas}$\tablenotemark{c}}&\colhead{SDSS spectrum} \\ \cline{4-5}
 & & & \colhead{inst+inst}& \colhead{inst+cont}& & &
}
\startdata
Mrk 94        & 310 & 1.29$\pm$0.02 & 1.29$\pm$0.11 & 1.33$\pm$0.17 & 8.07 &0.17  &0445-51873-404 \\
Mrk 1236      & 292 & 1.24$\pm$0.02 & 1.27$\pm$0.06 & 1.36$\pm$0.10 & 8.12 &0.14  &0481-51908-289 \\
J1402+5416    & 288 & 0.91$\pm$0.03 & 0.93$\pm$0.05 & 0.96$\pm$0.07 & 8.53 &0.17  &1323-52797-008 \\
J1404+5423    & 284 & 1.34$\pm$0.02 & 1.60$\pm$0.09 & 1.59$\pm$0.08 & 8.06 &0.01  &1325-52762-353 \\
J1044+0353    & 281 & 1.94$\pm$0.06 & 1.84$\pm$0.16 & 1.82$\pm$0.13 & 7.46 &0.06  &0578-52339-060 \\ 
CGCG 007-025  & 262 & 1.55$\pm$0.03 & 1.56$\pm$0.14 & 1.43$\pm$0.21 & 7.84 &0.00  &0266-51630-100 \\ 
HS 1734+5704  & 257 & 1.25$\pm$0.02 & 1.19$\pm$0.20 & 1.38$\pm$0.19 & 8.10 &0.03  &0358-51818-504 \\ 
J1403+5422    & 249 & 0.97$\pm$0.03 & 1.00$\pm$0.10 & 0.99$\pm$0.11 & 8.48 &0.08  &1325-52762-352 \\
J0248-0817    & 246 & 1.37$\pm$0.03 & 1.47$\pm$0.25 & 1.39$\pm$0.25 & 7.98 &0.14  &0456-51910-076 \\ 
J1402+5414    & 246 & 0.94$\pm$0.02 & 0.96$\pm$0.03 & 0.95$\pm$0.09 & 8.51 &0.16  &1323-52797-002 \\
J1253-0312    & 231 & 1.35$\pm$0.02 & 1.24$\pm$0.20 & 1.10$\pm$0.20 & 8.04 &0.08  &0337-51997-097 \\ 
HS 0837+4717  & 217 & 1.92$\pm$0.05 & 1.75$\pm$0.16 & 1.71$\pm$0.26 & 7.60 &0.32  &0550-51959-092 \\ 
J1304-0333    & 201 & 1.10$\pm$0.04 & 1.00$\pm$0.02 & 1.01$\pm$0.03 & 8.15 &0.50  &0339-51692-083 \\
J1228+4407    & 193 & 0.95$\pm$0.02 & 0.91$\pm$0.06 & 1.04$\pm$0.07 & 8.47 &0.26  &1371-52821-059 \\
Mrk 1318      & 169 & 1.01$\pm$0.04 & 1.05$\pm$0.12 & 1.05$\pm$0.17 & 8.35 &0.53  &0844-52378-041 \\ 
J1404+5425    & 163 & 1.10$\pm$0.03 & 0.93$\pm$0.05 & 0.92$\pm$0.11 & 8.46 &0.08  &1325-52762-350 \\
I Zw 18SE     & 147 & 1.72$\pm$0.07 & 1.66$\pm$0.42 & 1.78$\pm$0.41 & 7.25 &0.10  &0556-51991-312 \\ 
J1403+5421    & 138 & 0.86$\pm$0.13 & 0.80$\pm$0.08 & 0.74$\pm$0.11 & 8.17 &0.29  &1325-52762-345 \\
J1403+5418    & 128 & 0.93$\pm$0.17 & 0.77$\pm$0.10 & 0.80$\pm$0.11 & 8.24 &0.30  &1325-52762-356 \\
J1624-0022    & 108 & 1.17$\pm$0.03 & 1.11$\pm$0.21 & 1.24$\pm$0.18 & 8.20 &0.10  &0364-52000-187 \\ 
CGCG 032-017  & 108 & 1.40$\pm$0.03 & 1.31$\pm$0.15 & 1.40$\pm$0.23 & 8.08 &0.40  &1185-52642-123 \\ 
SBS 1222+614  &  93 & 1.38$\pm$0.03 & 1.38$\pm$0.25 & 1.35$\pm$0.27 & 8.00 &0.04  &0955-52409-608 \\ 
J1038+5330    &  93 & 0.98$\pm$0.03 & 0.95$\pm$0.14 & 0.92$\pm$0.16 & 8.37 &0.14  &1010-52649-328 \\
J1205+5032    &  72 & 0.84$\pm$0.28 & 0.88$\pm$0.13 & 0.84$\pm$0.19 & 8.25 &0.62  &0969-52442-489 \\
\enddata
\tablenotetext{a}{$t_e$(O {\sc iii}) = 
$T_e$(O {\sc iii})/10000 K, derived from the [O {\sc iii}] $\lambda$(4959+5007)/$\lambda$4363 emission line
ratio.}
\tablenotetext{b}{$t_e$(H$^+$) = $T_e$(H$^+$)/10000, derived as the 
weighted mean temperature from the 10 best fits of the SED.}
\tablenotetext{c}{Equivalent widths of the H$\beta$ emission line, 
oxygen abundances and extinction coefficients derived from the 
observed Balmer decrement are taken from \citet{I05}.}
\end{deluxetable}

\clearpage

\begin{figure*}
\figurenum{1}
\epsscale{0.9}
\plotone{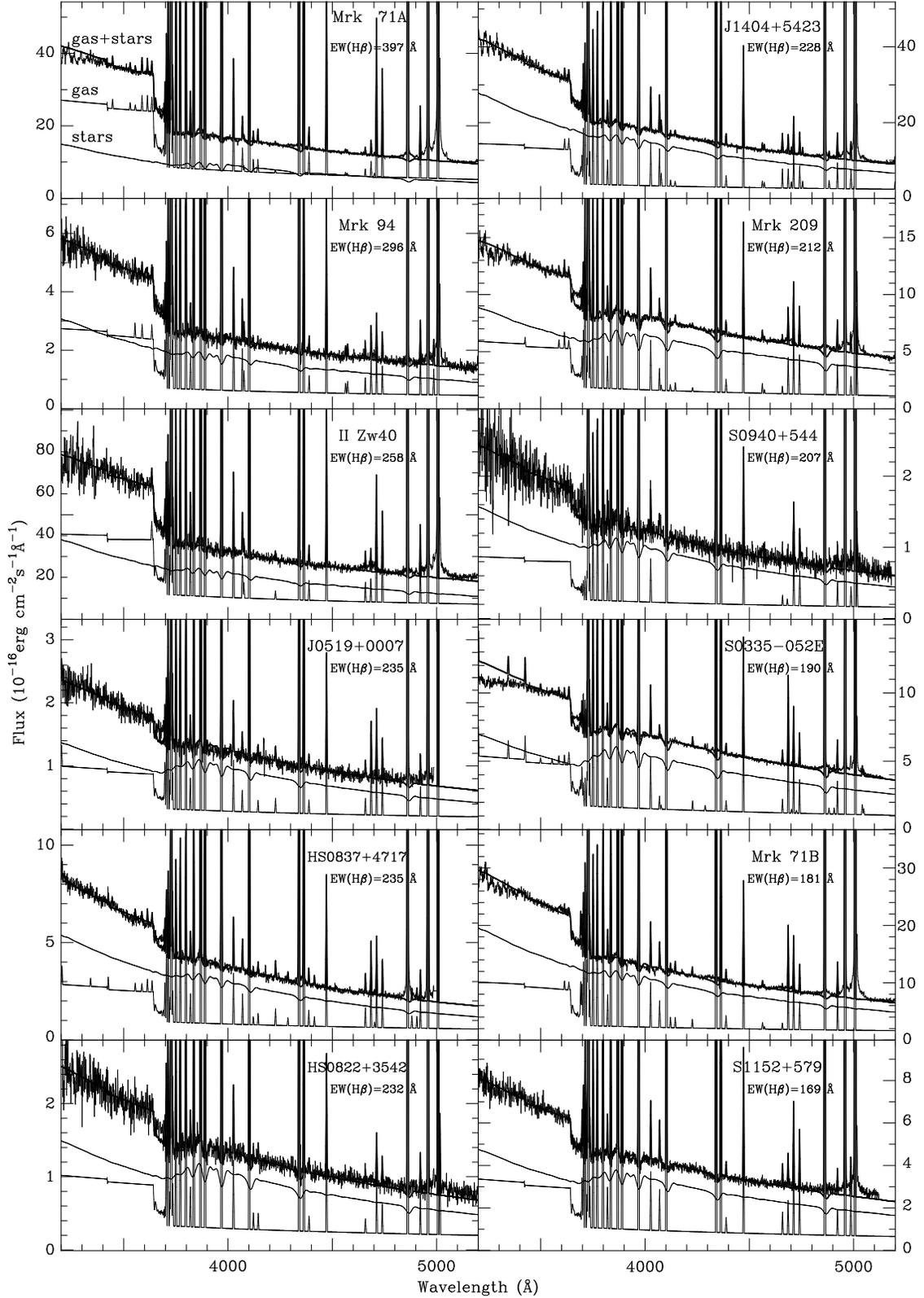}
\caption{Best fit model SEDs (thick solid lines) 
superposed by the redshift- and extinction-corrected MMT  
spectra of 22 H {\sc ii} regions in 18 blue compact dwarf galaxies and 
in one
H {\sc ii} region J1404+5423 of the spiral galaxy M101. 
The separate contributions 
from the stellar and ionized gas components are shown
by thin solid lines (see the labels in the upper left panel for Mrk 71A).   
\label{sp_1}}
\end{figure*}

\clearpage

\begin{figure*}
\figurenum{1}
\epsscale{0.9}
\plotone{f1b.eps}
\caption{Continued.
    \label{sp_2}}
\end{figure*}

\clearpage

\begin{figure*}
\figurenum{2}
\epsscale{1.0}
\plotone{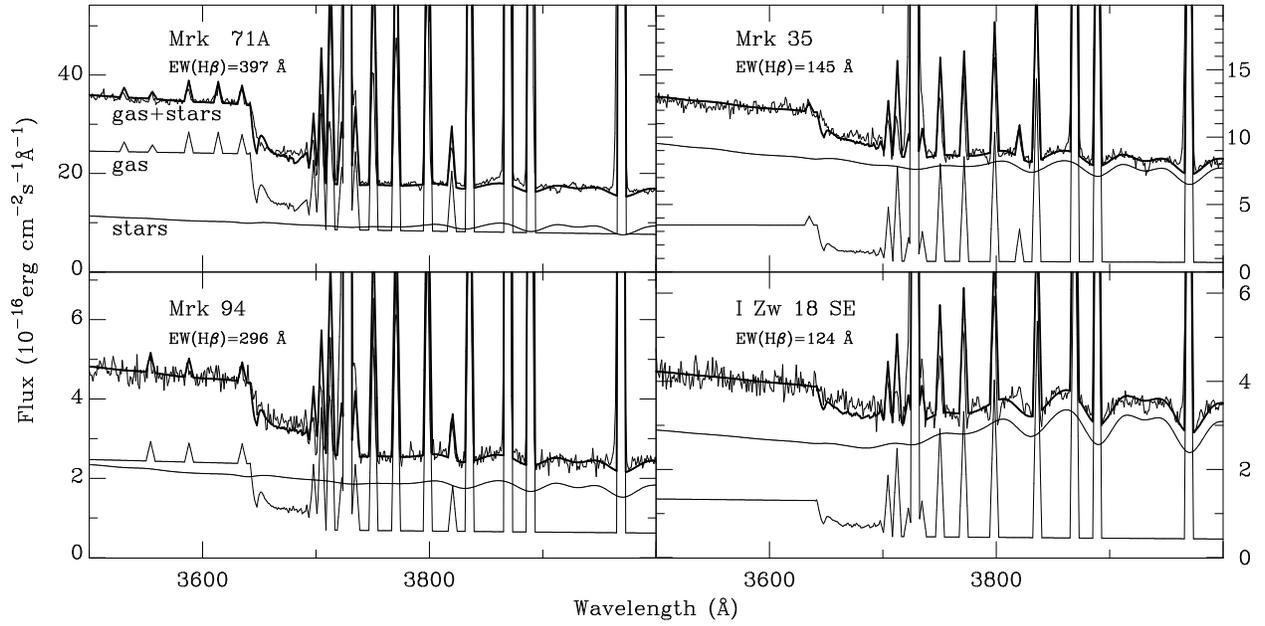}
\caption{Blow-up of the Balmer jump region in 4 selected spectra of  
Fig. \ref{sp_1}.
    \label{sp_3}}
\end{figure*}

\clearpage

\begin{figure*}
\figurenum{3}
\epsscale{0.9}
\plotone{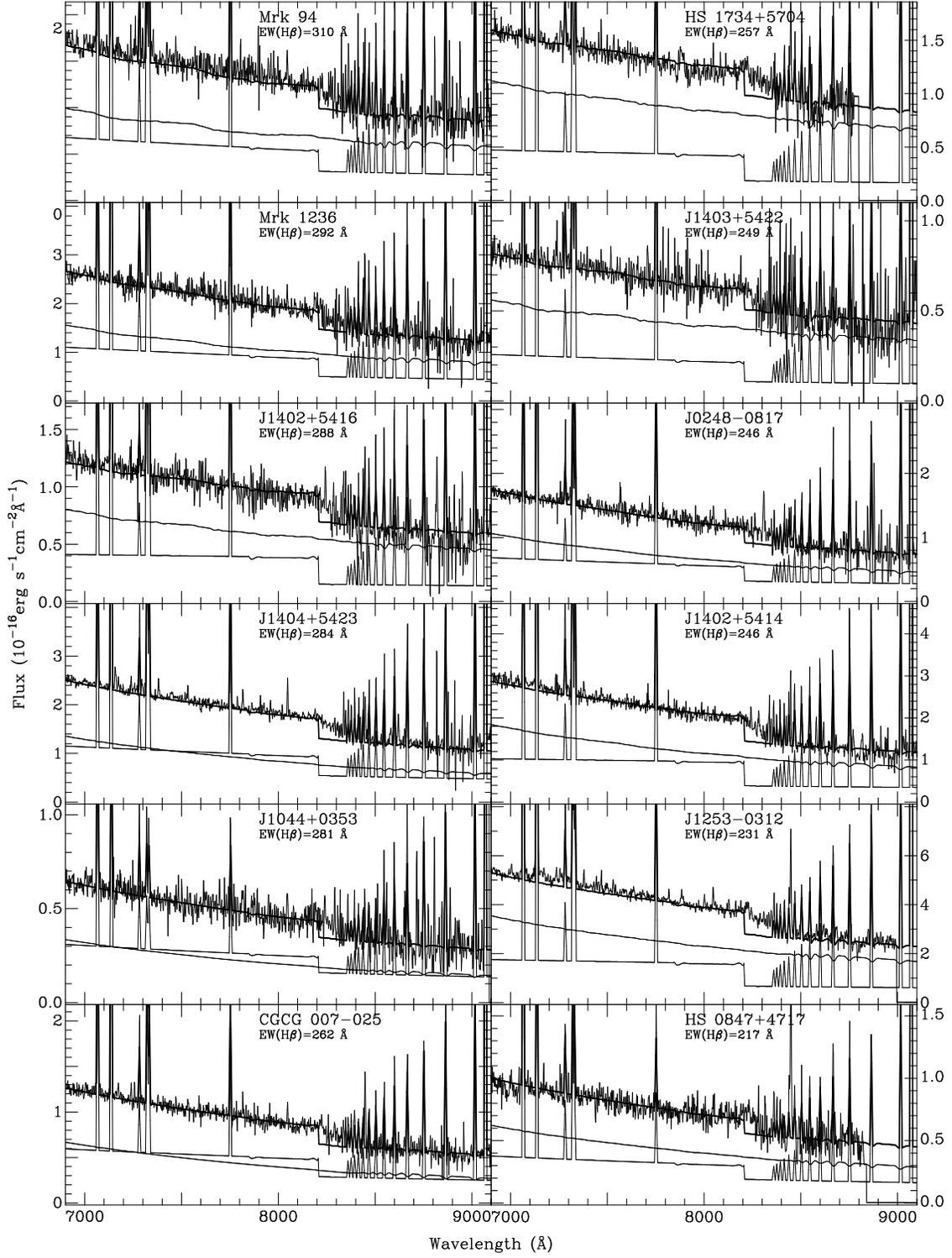}
\caption{Best fit model SEDs (thick solid lines) 
superposed on the red parts of the redshift- and extinction-corrected SDSS  
spectra of 11 H {\sc ii} regions in blue compact dwarf galaxies and 
of 13 H {\sc ii} regions in spiral galaxies. 
The separate contributions 
from the stellar and ionized gas components are shown
by thin solid lines.   
\label{spp_1}}
\end{figure*}

\clearpage

\begin{figure*}
\figurenum{3}
\epsscale{0.9}
\plotone{f3b.eps}
\caption{Continued.
    \label{spp_2}}
\end{figure*}

\clearpage

\begin{figure*}
\figurenum{4}
\epsscale{1.1}
\plottwo{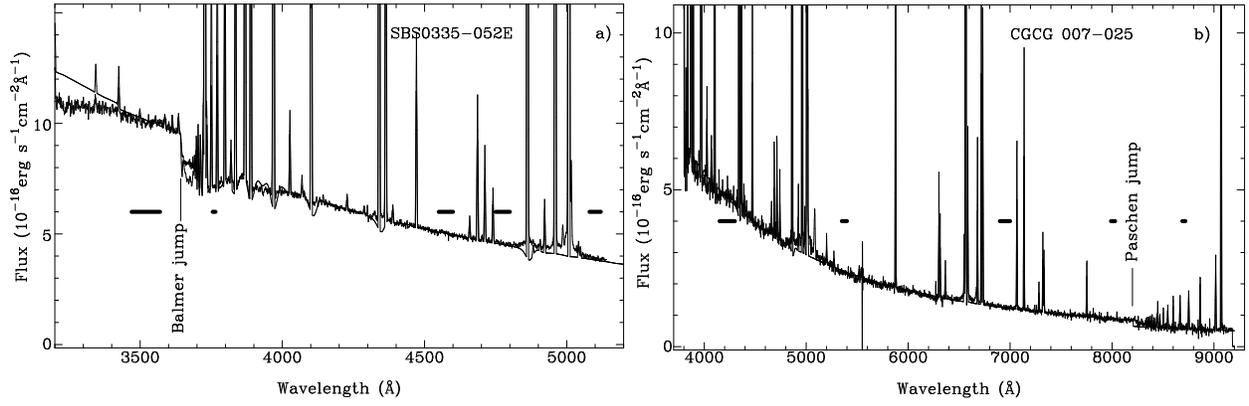}{f4b.eps}
\caption{Best fit model SED (thick solid lines) superposed on 
the redshift- and extinction-corrected 
spectra of SBS 0335--052E (a) and CGCG 007--025 (b). Locations of the Balmer
jump in (a) and of the Paschen jump in (b) are marked by vertical lines. 
The five selected continuum regions used for fitting the observed spectrum
are shown by thick horizontal lines in each panel.
\label{spcgcg}}
\end{figure*}

  \begin{figure}
\figurenum{5}
\epsscale{0.4}
\plotone{f5a.eps}
\plotone{f5b.eps}
\plotone{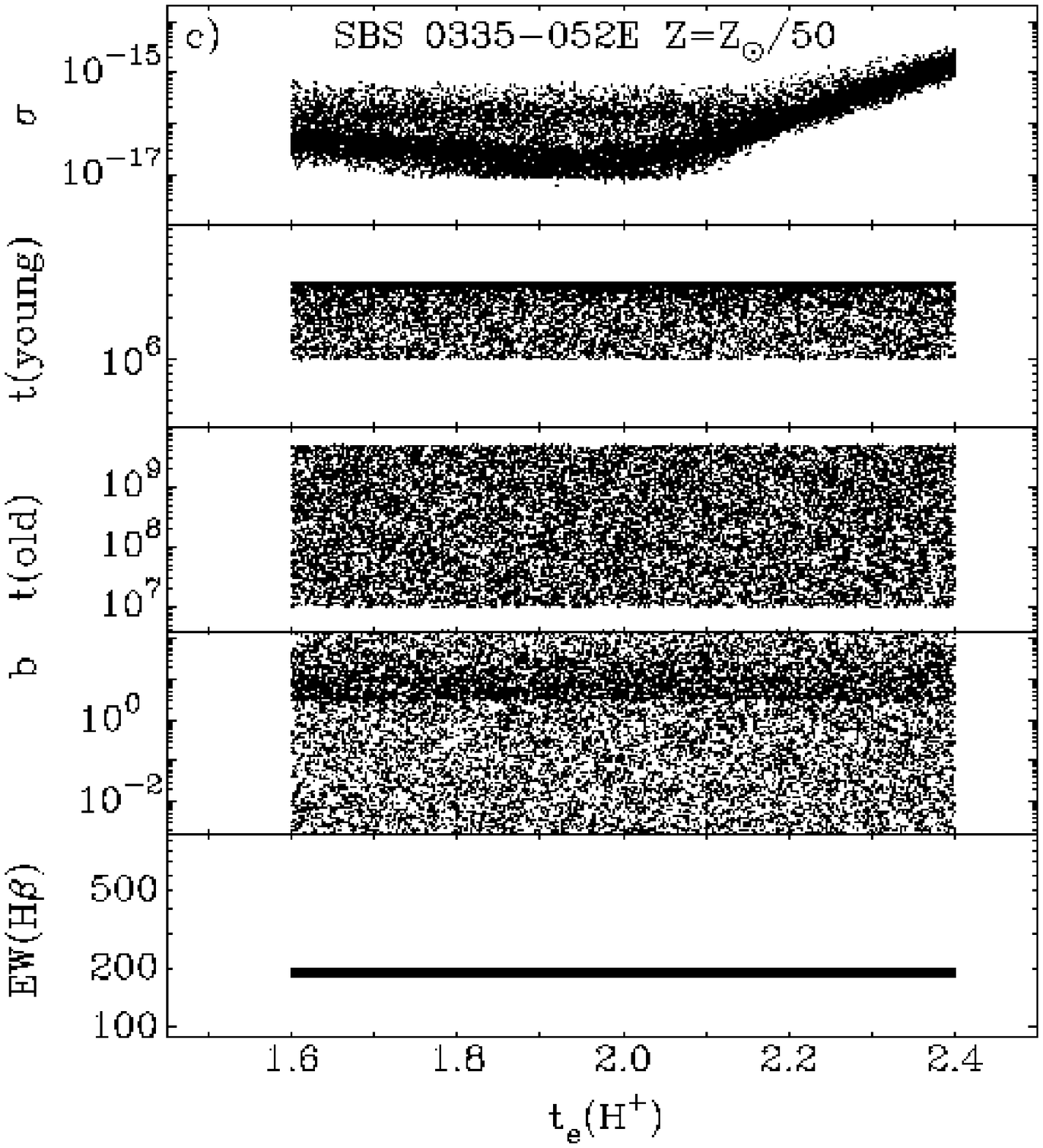}
\plotone{f5d.eps}
    \caption{Parameter space explored in Monte Carlo simulations to fit the 
spectra of two BCDs: one is Mrk 35 (panel a),  
the highest-metallicity BCD in our sample and the other is 
SBS 0335--052E (panel c), one of the 
lowest-metallicity BCD in the sample. The parameters shown are: 
the equivalent width EW(H$\beta$), 
the ages of the old and young stellar populations, $t$(old) and $t$(young),
the mass ratio $b$ of the old-to-young stellar populations,
and the electron temperature $t_e$(H$^+$) in the H$^+$ zone.
$\sigma$ is an estimator of the goodness of the fit.
Panels (b) and (d) are the same as panels (a) and (c) except that only 
the 10 best Monte Carlo realizations are shown.
    \label{dmetal}}
\end{figure}

\clearpage

\begin{figure}
\figurenum{6}
\epsscale{0.5}
\plotone{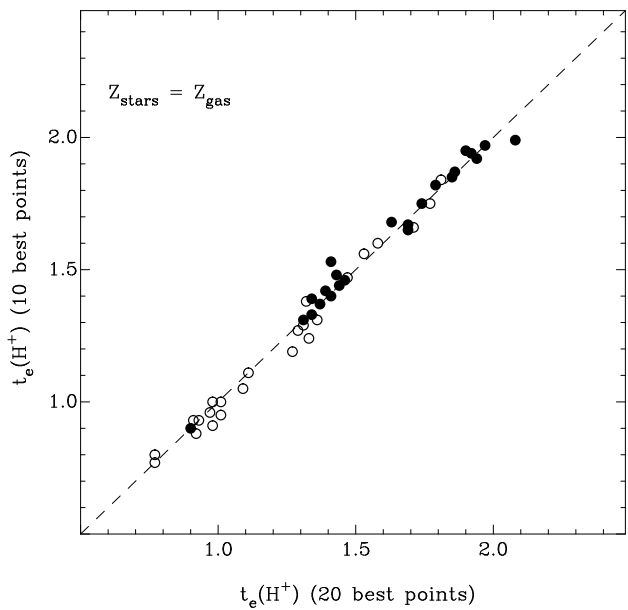}
\caption{The weighted mean temperature $t_e$(H$^+$) of the 10 best SED fits
among all Monte Carlo realizations vs. the weighted mean
$t_e$(H$^+$) of the 20 best SED fits among all Monte Carlo realizations
for each BCD. Equality is shown by a dashed line. H {\sc ii} regions
observed with the MMT are shown by filled circles and those 
from the SDSS by open circles.
    \label{t_10_20}}
\end{figure}

\clearpage

\begin{figure}
\figurenum{7}
\epsscale{0.5}
\plotone{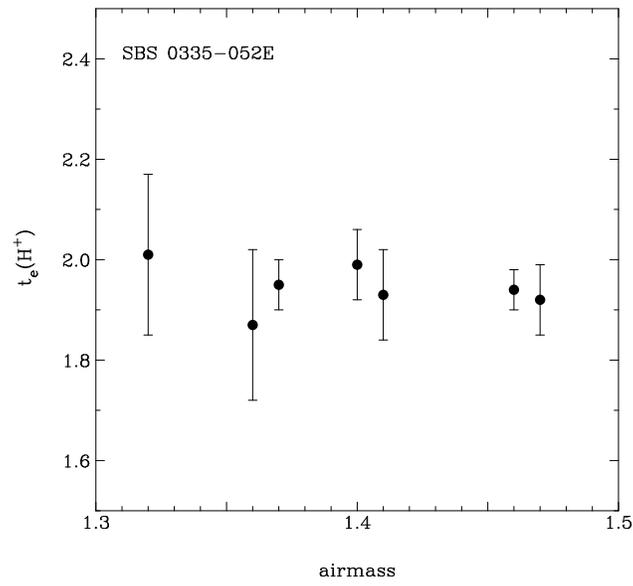}
\caption{The weighted mean temperature $t_e$(H$^+$) derived
from different spectra of SBS 0335--052E as a function of airmass.
    \label{atmref}}
\end{figure}

\clearpage

\begin{figure}
\figurenum{8}
\epsscale{0.5}
\plotone{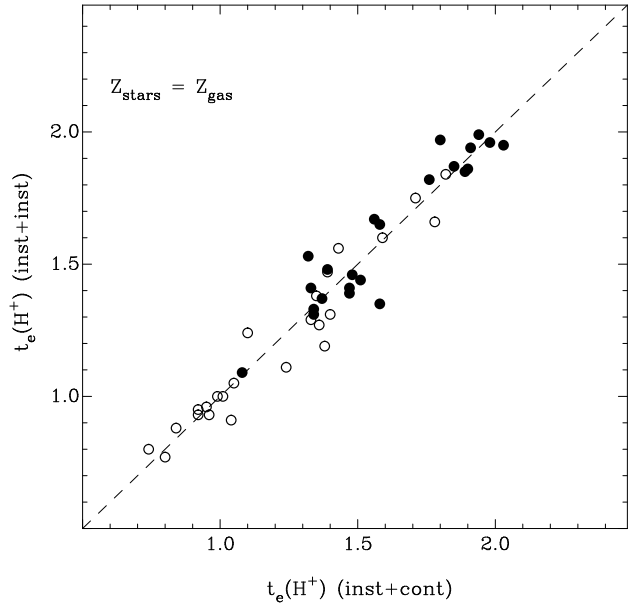}
\caption{Dependence of the derived temperatures in the H+ regions on the 
star formation history.
The quantity $t_e$(H$^+$) (inst+inst) denotes the temperature derived by 
fitting the galaxy's spectrum with a two-burst model, one for the young 
stellar population and one for the old stellar population, while the 
quantity $t_e$(H$^+$) (inst+cont)
denotes the temperature derived by fitting the galaxy's spectrum
with a model where the young stellar population is made in a
short burst and the old stellar population in a continuous mode.
No systematic difference is seen between the two temperatures.
    \label{t_inst_cont}}
\end{figure}

\clearpage

\begin{figure}
\figurenum{9}
\epsscale{0.7}
\plotone{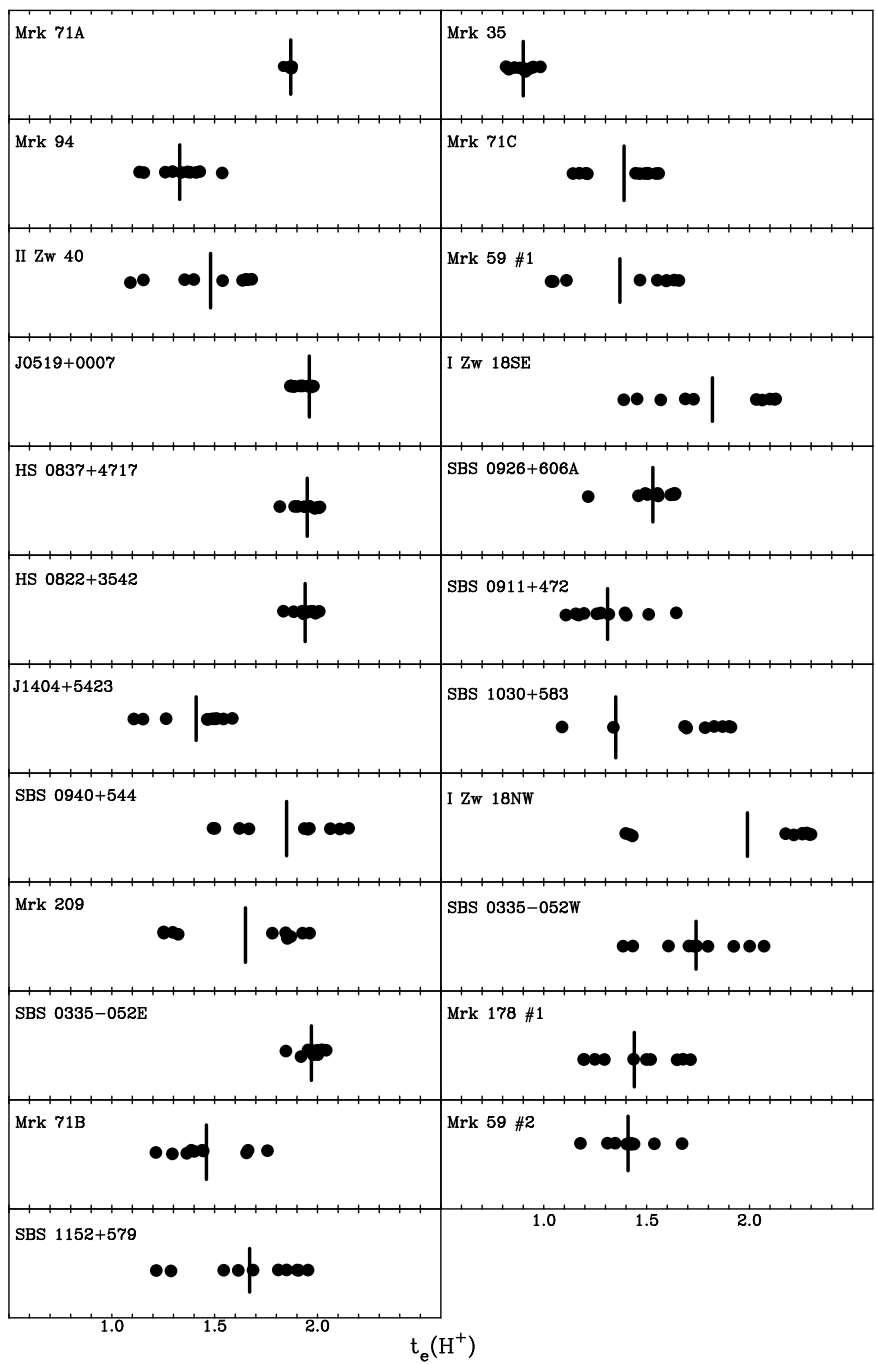}
\caption{Temperature $t_e$(H$^+$) = 10$^{-4}$$T_e$(H$^+$) of the 10 best 
SED fits (filled circles) among all Monte Carlo realizations
for each H {\sc ii} region observed with the MMT. 
The mean temperatures weighted by $\sigma$ are shown by straight
vertical lines.
    \label{t_e}}
\end{figure}

\clearpage

\begin{figure}
\figurenum{10}
\epsscale{0.7}
\plotone{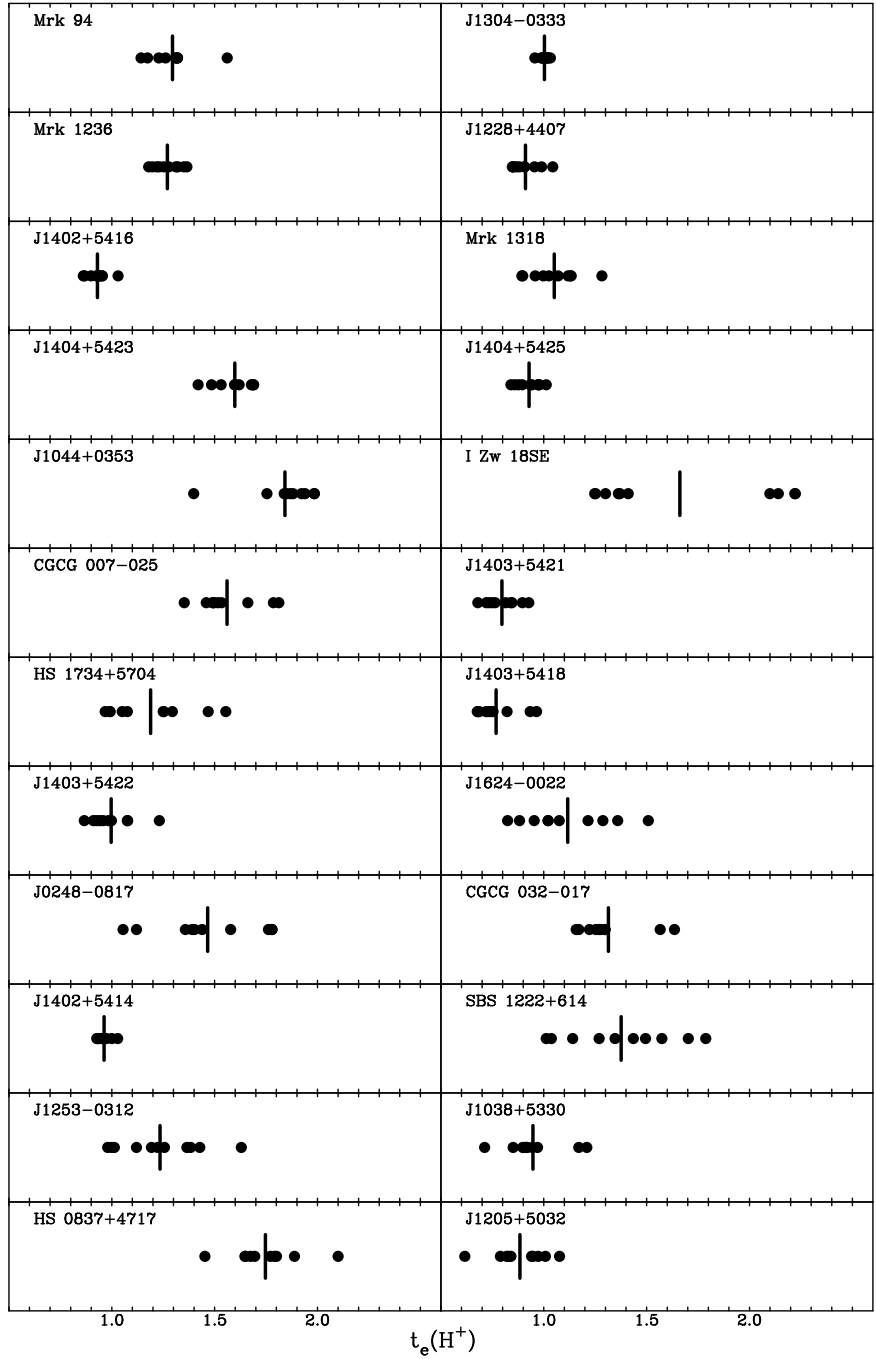}
\caption{Temperature $t_e$(H$^+$) = 10$^{-4}$$T_e$(H$^+$) of the 10 best 
SED fits (filled circles) among all Monte Carlo realizations
for each H {\sc ii} region extracted from the SDSS. 
The mean temperatures weighted by $\sigma$ are shown by straight
vertical lines.
    \label{t_ep}}
\end{figure}

\clearpage

\begin{figure}
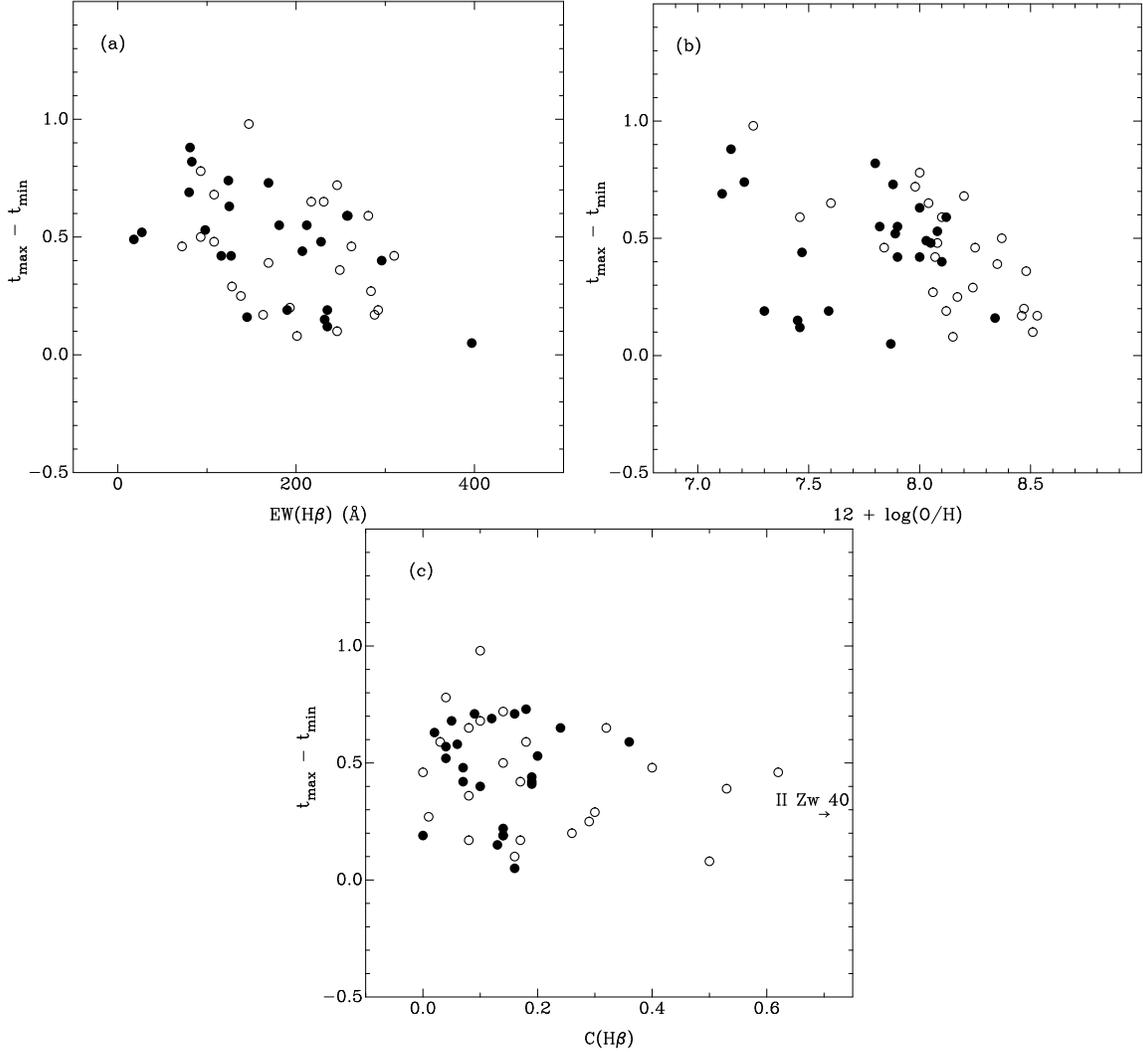

\figurenum{11}
\epsscale{0.45}
\plotone{f11a.eps}
\plotone{f11b.eps}
\plotone{f11c.eps}
\caption{Maximal differences ($t_{max}$ -- $t_{min}$) of the temperature 
$t_e$(H$^+$) among the 10 best Monte Carlo realizations
in each galaxy vs (a) the equivalent width of 
H$\beta$, (b)  the oxygen abundance 12 + log(O/H) and (c) 
the extinction coefficient
$C$(H$\beta$). In all panels, the metallicity of the stellar population is set 
equal to that of the stellar evolutionary model 
closest to the metallicity of the gas. In panel (c), the extinction of 
the BCD II Zw 40 is off-scale because it is very nearly in the Galactic plane 
and is heavily reddened by Galactic extinction. Filled circles show 
H {\sc ii} regions observed with the MMT, and open circles those 
from the SDSS.
\label{delta_t}}
\end{figure}

\clearpage

  \begin{figure}[hbtp]
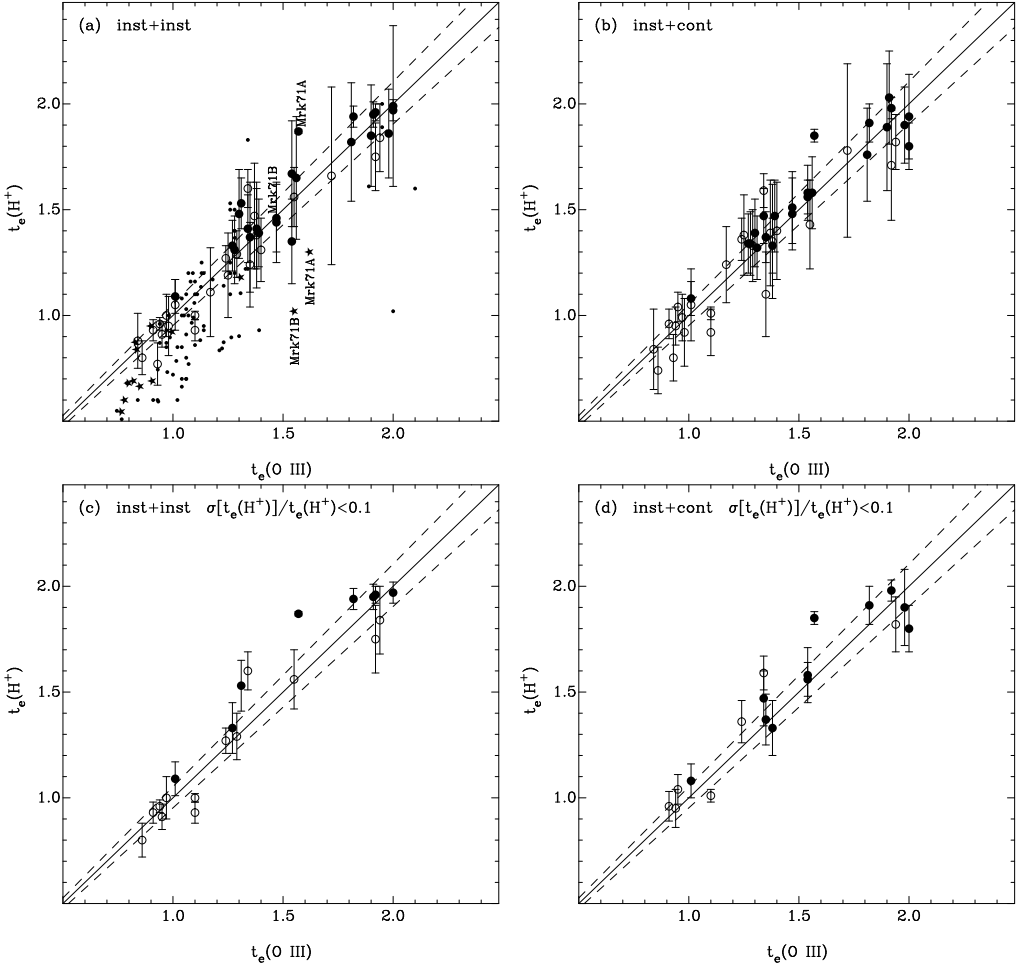

\figurenum{12}
\epsscale{0.4}
\plotone{f12a.eps}
\plotone{f12b.eps}
\plotone{f12c.eps}
\plotone{f12d.eps}
    \caption{
Comparison of the temperature 
$t_e$(H$^+$) derived from fitting the Balmer jump and the 
SED and of the temperature $t_e$(O {\sc iii}) derived 
from the nebular to auroral line 
intensity ratio of doubly ionized 
oxygen [O {\sc iii}] $\lambda$(4959+5007)/$\lambda$4363, for all objects in 
our sample (filled circles show H {\sc ii} regions observed with the MMT
and open circles those from the SDSS). 
Two cases have been considered: 
 (a) the two-burst (inst+inst) model, and (b) 
the burst+continuous star formation (inst+cont) model.
Panels (c) and (d) are the same as panels (a)
and (b), but show only H {\sc ii} regions with deep minima for $t_e$(H$^+$).
In each panel, the solid line denotes equal temperatures  
and the dashed lines show a $\pm$5\% deviation
from the line of equal temperatures. 
Error bars are the root mean square temperature deviations from the 
mean of the 10 best 
 Monte Carlo simulations. 
Stars in (a) show the data for H {\sc ii} regions
from \citet{E98}, \citet{G04,G06}, \citet{P92}, \citet{P00}, \citet{P03} and 
\citet{G94}. For comparison, we have labeled the Mrk 71 A, B 
data points from the present work and from \citet{G94}. 
Our $t_e$(H$^+$) for Mrk 71 A, B derived using the Balmer jump are 
considerably 
higher than those of \citet{G94} using the Paschen jump. 
For comparison, we have also plotted in (a) 
the data of \citet{Liu2004}, \citet{Z2004} and \citet{W05}
for 54 Galactic PNe (dots).
 \label{tP0}}
\end{figure}

\end{document}